\documentclass[a4paper,11pt]{article}
\pdfoutput=1
\usepackage{jcap}
\usepackage{amsmath,amssymb,bm,float,Macro,mathrsfs,tabularx,url}
\usepackage{threeparttable}
\def\tn#1{#1}
\def\hpol{\widehat{p}}
\def\estk{\widehat{\kappa}}
\def\paperI{LB-Lensing}
\def\pteppaper{LB23}
\def\LB{{\it LiteBIRD}}
\def\Planck{{\it Planck}}

\def\Euclid{{\it Euclid}}
\def\S4{CMB-S4}

\author[1]{T.\,Namikawa,}
\author[2,3,20]{A.\,I.\,Lonappan,}
\author[3,4,5]{C.\,Baccigalupi,}
\author[7,8,9]{N.\,Bartolo,}
\author[10]{D.\,Beck,}
\author[11]{K.\,Benabed,}
\author[12,13,14]{A.\,Challinor,}
\author[6,15]{P.\,Diego-Palazuelos,}
\author[16]{J.\,Errard,}
\author[17]{S.\,Farrens,}
% ...
\affiliation[1]{Kavli Institute for the Physics and Mathematics of the Universe (Kavli IPMU, WPI), UTIAS, The University of Tokyo, Kashiwa, Chiba 277-8583, Japan}
\affiliation[2]{Dipartimento di Fisica, Università di Roma Tor Vergata, Via della Ricerca Scientifica, 1, 00133, Roma, Italy}
\affiliation[3]{International School for Advanced Studies (SISSA), Via Bonomea 265, 34136, Trieste, Italy}
\affiliation[4]{INFN Sezione di Trieste, via Valerio 2, 34127 Trieste, Italy}
\affiliation[5]{IFPU, Via Beirut, 2, 34151 Grignano, Trieste, Italy}
\affiliation[6]{Instituto de Fisica de Cantabria (IFCA, CSIC-UC), Avenida los Castros SN, 39005, Santander, Spain}
\affiliation[7]{Dipartimento di Fisica e Astronomia “G. Galilei”, Universita` degli Studi di Padova, via Marzolo 8, I-35131 Padova, Italy}
\affiliation[8]{INFN Sezione di Padova, via Marzolo 8, I-35131, Padova, Italy}
\affiliation[9]{INAF, Osservatorio Astronomico di Padova, Vicolo dell’Osservatorio 5, I-35122, Padova, Italy}
\affiliation[10]{Stanford University, Department of Physics,  CA 94305-4060, USA}
\affiliation[11]{Institut d'Astrophysique de Paris, CNRS/Sorbonne Université, Paris France}
\affiliation[12]{DAMTP, Centre for Mathematical Sciences, Wilberforce Road, Cambridge CB3 0WA, U.K.}
\affiliation[13]{Institute of Astronomy, Madingley Road, Cambridge CB3 0HA, U.K.}
\affiliation[14]{Kavli Institute for Cosmology Cambridge, Madingley Road, Cambridge CB3 0HA, U.K.}
\affiliation[15]{Dpto. de Física Moderna, Universidad de Cantabria, Avda. los Castros s/n, E-39005 Santander, Spain}
\affiliation[16]{Université de Paris, CNRS, Astroparticule et Cosmologie, F-75013 Paris, France}
\affiliation[17]{AIM, CEA, CNRS, Université Paris-Saclay, Université de Paris, F-91191 Gif-sur-Yvette, France}
% ...
\affiliation[20]{INFN Sezione di Roma2, Università di Roma Tor Vergata, via della Ricerca Scientifica, 1, 00133 Roma, Italy}
% ...

\emailAdd{toshiya.namikawa@ipmu.jp}

\abstract{
We estimate the efficiency of mitigating the lensing $B$-mode polarization, the so-called delensing, for the \LB\ experiment with multiple external data sets of lensing-mass tracers. 
The current best bound on the tensor-to-scalar ratio, $r$, is limited by lensing rather than Galactic foregrounds. 
Delensing will be a critical step to improve sensitivity to $r$ as measurements of $r$ become more and more limited by lensing. 
In this paper, we extend the analysis of the recent \LB\ forecast paper to include multiple mass tracers, i.e., the CMB lensing maps from \LB\ and \S4-like experiment, cosmic infrared background, and galaxy number density from \Euclid- and LSST-like survey. 
We find that multi-tracer delensing will further improve the constraint on $r$ by about $20\%$. In \LB, the residual Galactic foregrounds also significantly contribute to uncertainties of the $B$-modes, and delensing becomes more important if the residual foregrounds are further reduced by an improved component separation method. 
}

\title{
LiteBIRD Science Goals and Forecasts: Improving Sensitivity to Inflationary Gravitational Waves with Multitracer Delensing
}

\begin{document}
\maketitle
\flushbottom

\section{Introduction} 

Measuring the polarization of the cosmic microwave background (CMB) anisotropies will be at the forefront of observational cosmology in the next decade. In particular, measurements of the curl component ($B$ modes) in the CMB polarization will be of great importance, as these provide us with a unique window to probe inflationary gravitational waves (IGWs) and gain new insights into the early Universe~\cite{Kamionkowski:1996:GW,Seljak:1996:GW} (see also reviews \cite{Kamionkowski:2015yta,Komatsu:2022:review}). 
%CMB observations have not yet confirmed the presence of these IGWs but have placed upper bounds on the IGW amplitude parameterized by the tensor-to-scalar ratio $r$. The recent data from the BICEP/Keck Array,  combined with \Planck\ and \WMAP\ data, places the constraint on $r$ at a pivot scale of $0.05\,$Mpc$^{-1}$ as $r<0.036$ ($2\,\sigma$)~\cite{BK13:GW,BK15:data}.
The BICEP/Keck Array collaborations recently placed constraints on the IGW background, parameterized by the tensor-to-scalar ratio $r$ (at a pivot scale of $0.05\,$Mpc$^{-1}$) as $r<0.036$ ($2\,\sigma$) for a fixed cosmology \cite{BK13:GW,BK15:data}, while Ref.~\cite{Tristram:2021:r} combine the BICEP/Keck Array data with \Planck\ PR4 and baryon acoustic oscillations, finding $r<0.032$ ($2\,\sigma$) by simultaneously constraining the cosmological parameters.
Several ongoing and upcoming CMB experiments, including the BICEP Array \cite{BICEPArray}, Simons Array \cite{SimonsArray}, Simons Observatory (SO) \cite{SO:2018:forecast}, \LB\ \cite{LiteBIRD:PTEP}, and \S4\ \cite{CMBS4:r-forecast}, are targeting detection of IGW $B$-mode polarization over the next decade.

A high-precision measurement of the large-scale $B$-mode polarization can tightly constrain $r$. However, the precision of the IGW $B$-mode measurement is limited by other sources of $B$ modes. On large angular scales, the Galactic foregrounds dominate the IGW $B$-mode polarization \cite{B2I}. To overcome this issue, many studies have developed some methodology to substantially reduce the Galactic foreground contaminants (e.g., Refs.~\cite{Delabrouille:2002:SMICA,WMAP:2003:Bennett,Tegmark:2003:HILC,Eriksen:2004:FG-clean,Eriksen:2007:Commander,Delabrouille:2008qd,Stompor:2008:FGbuster,Katayama:2011eh,Errard:2016:FG,Ichiki:2018:delta-map,Remazeilles:2020:cMILC,Carones:2022:FG}, and Ref.~\cite{LiteBIRD:PTEP} for the \LB\ case). In addition to Galactic foregrounds, gravitational lensing leads to $B$ modes from the conversion of part of the $E$ modes \cite{Zaldarriaga:1998:LensB}; this lensing $B$-mode polarization is as an additional noise component when constraining $r$. 
The lensing $B$-mode polarization on large scales behaves as white noise, and its amplitude is comparable to that of the IGW $B$-mode polarization with $r=0.01$ at the recombination bump ($\ell\sim 10$ -- $100$) \cite{Lewis:2006:review}. 
The constraint on $r$ is already limited by the variance from the lensing $B$-mode polarization more than by Galactic foregrounds, at least, in low dust region \cite{BK13:GW}. 
Reducing statistical uncertainties by subtracting off the lensing $B$-mode polarization or equivalent methods -- a process usually referred to as ``delensing'' -- will hence be critical for improving the constraints on $r$ (e.g., Refs.~\cite{Kesden:2002ku,Seljak:2003:delens,Millea:2017:bayesian,Carron:2018:GW,Caldeira:2018ojb,Millea:2020:delens,Hotinli:2021:delens,Heinrich:2022hsf,Aurlien:2022tlp,Yan:2023oan,Ni:2023ume,Belkner:2023:s4delens,Trendafilova:2023xtq}). 

To estimate the lensing $B$-mode polarization in some survey region, known as a lensing $B$-mode template, the simplest method is to combine the measured $E$-mode polarization with a reconstructed lensing map derived from CMB data, and multiple works have studied this technique (e.g., Refs.~\cite{Seljak:2003:delens,Smith:2010:delens,Teng:2011:delens,Hanson:2013:lensb,Namikawa:2014:patchwork,Sehgal:2016,P16:lensB,Namikawa:2017:delens,Carron:2017:Planck-delens,SPTpol:delens,PB:2019:delens,BaleatoLizancos:2020:cibdelens,Beck:2020:FG-lens,ACT:DW:2020,Reinecke:2023gtp}). In addition, other mass tracers, such as the spatial distribution of optical and radio galaxies, as well as the cosmic infrared background (CIB) are also possible to include measurement of the lensing $B$-mode polarization (e.g., Refs.~\cite{Sigurdson:2005:delens,Marian:2007,Simard:2015,Sherwin:2015,Namikawa:2015:delens,Challinor:2018:core-lens,Manzotti:2018:delens}). Recently, Ref.~\cite{BKSPT} demonstrated for the first time $B$-mode delensing to improve constraints on IGWs using the CIB as a mass tracer. In the \LB\ case, the CIB and galaxies are useful for delensing. 
In addition, the lensing potential reconstructed from \S4\ data will be available in the \S4\ observing patch. 

This work is part of a series of papers that present the science achievable by the \LB\ space mission, expanding on the overview published in Ref.~\cite{LiteBIRD:PTEP} (hereafter, \pteppaper). 
In particular, this work focuses on the potential of the \LB\ $B$-mode delensing to improve sensitivity to constrain IGWs by combining multiple external mass tracers. 
In \pteppaper, we consider a simple forecast using CIB as a possible mass tracer for delensing. In this paper, we extend the delensing study in several ways. We include the CMB lensing maps of \LB\ and \S4, the CIB, and galaxy number density from \Euclid\ and the Vera Rubin Observatory Legacy Survey of Space and Time (LSST). 
In addition, we make our forecast more realistic by adding survey windows of each survey, a realistic lensing map from \LB, component separation of the \LB\ $E$-mode polarization used for the lensing $B$-mode template, and the large-scale $B$-mode polarization after the component separation obtained in \pteppaper. 
\tn{
We also note that \pteppaper\ provides the delensing forecast only for the case if the true value of $r$ is $r_0=0$. If $r_0$ is $\mC{O}(10^{-3})$, the cosmic variance from IGWs significantly dominates the statistical uncertainty at the reionization bump but does not at the recombination bump. On the other hand, if $r_0$ goes to zero, the information from the reionization bump also becomes important to constrain $r$. Since the lensing cosmic variance is important only at the recombination bump, delensing would become more important for $r_0=\mC{O}(10^{-3})$ than the case with $r_0=0$. Also, if $r_0$ is close to or less than the \LB\ target sensitivity, $10^{-3}$, the analysis needs an accurate treatment of possible concerns, such as instrumental systematic effects and foreground contaminations, whose characterization is beyond the scope of this paper. Thus, this paper explores the case for a nonzero $r_0$.
}

This paper is organized as follows. 
Section~\ref{sec:multitracer} presents the baseline strategy for \LB\ delensing after briefly reviewing the lensing effect on the CMB. 
Section~\ref{sec:simulation} describes our simulation setup. 
Section~\ref{sec:results} shows the improvement in constraint on $r$ by the multi-tracer delensing, including realistic survey effects.
Section~\ref{sec:summary} is devoted to summary and discussion. 
This paper is a companion paper of the \LB\ lensing study \cite{LiteBIRD:lens} (hereafter, \paperI) and uses some of the results presented in that paper.

\section{Multitracer delensing} \label{sec:multitracer}

\subsection{Lensing-induced {\it B}-modes and delensing}

The distortion effect of lensing on the primary CMB polarization anisotropies is expressed by remapping the primary polarization anisotropies from the last-scattering surface, $Q\pm\iu U$. The lensed Stokes parameters in sky direction $\hatn$ can be approximately expressed as a remapping \cite{Lewis:2006:review}:
%--------------------------------------------------%
\al{
	[\tQ\pm\iu\tU](\hatn) &= [Q\pm\iu U](\hatn + \bm{d}(\hatn)) 
	\,, \label{Eq:remap}
}
%--------------------------------------------------%
where tildes indicate lensed quantities and $\bm{d}$ is the deflection angle used to describe the lensing displacements on the spherical sky (see Ref.~\cite{Challinor:2002cd} for more details). The quantity $\bm{d}$ is related to the lensing convergence, $\kappa$, as $\bn\cdot\bm{d}=-2\kappa$, where $\bn$ is the covariant derivative on the unit sphere. The $E$- and $B$-mode polarization are then obtained by \cite{Zaldarriaga:1996:EBdef,Kamionkowski:1996:eb}
%--------------------------------------------------%
\al{
	E_{\l m} \pm \iu B_{\l m} = -\Int{2}{\hatn}{} {}_{\pm 2}Y_{\l m}^*(\hatn) [Q\pm\iu U](\hatn) 
	\,, \label{Eq:EB-def}
}
%--------------------------------------------------% 
where we denote the spin-$2$ spherical harmonics as ${}_{\pm 2}Y_{\l m}(\hatn)$. Similarly with the spin-$0$ (scalar) spherical harmonics, $Y_{\l m}(\hatn)$, the harmonic coefficients of the lensing convergence is defined as
%--------------------------------------------------% 
\al{
	\kappa_{LM} &= \Int{2}{\hatn}{} Y_{LM}^*(\hatn) \kappa (\hatn) 
	\,.
}
%--------------------------------------------------% 
Expanding the right-hand side of \eq{Eq:remap} to first order in the
lensing potential and transforming the Stokes $Q/U$ parameters
to $E/B$ modes using \eq{Eq:EB-def}, the $B$ modes of the lensed polarization field can be expressed as \cite{Smith:2010:delens}:
%--------------------------------------------------%
\al{
    B^{\mathrm{lens}}_{lm} &= \iu\sum_{\l'm'}\sum_{LM} 
	\Wjm{\l}{\l'}{L}{m}{m'}{M} p^- F^{(2)}_{\l L\l'} E_{\l'm'}^* \kappa_{LM}^* 
	\label{Eq:Lensed-B} \,, 
} 
%--------------------------------------------------%
where we ignore the primary $B$-mode polarization. The quantity in round brackets is the Wigner-3$j$ symbol, $p^-$ is unity if $\l+L+\l'$ is an odd integer and zero otherwise, and $F^{(2)}_{\l L\l'}$ represents the mode coupling induced by the lensing \cite{OkamotoHu:quad}: 
%--------------------------------------------------%
\al{ 
	F^{(s)}_{\l L\l'} &= \frac{2}{L(L+1)}
	\sqrt{\frac{(2\l+1)(2\l'+1)(2L+1)}{16\pi}}
	\notag \\
	&\times 
	[-\l(\l+1)+\l'(\l'+1)+L(L+1)]\Wjm{\l}{\l'}{L}{-s}{s}{0} 
	\,. \label{Eq:Spm}
} 
%--------------------------------------------------%
From \eq{Eq:Lensed-B}, the lensing $B$-mode polarization is simply expressed in terms of a convolution between the unlensed $E$-mode polarization and the lensing potential. Once we obtain an estimate for the $E$ mode and lensing maps, we simply approximate the lensing $B$-mode polarization as a convolution of the Wiener-filtered $E$-mode polarization and lensing convergence:
%--------------------------------------------------%
\al{
	\widehat{B}^{\mathrm{temp}}_{\l m} = \iu\sum_{\l'm'}\sum_{LM} 
	\Wjm{\l}{\l'}{L}{m}{m'}{M}p^- F^{(2)}_{\l L\l'} 
	(\hE^{\rm WF}_{\l'm'})^*(\estk^{\rm WF}_{LM})^*
	\,, \label{Eq:Quad-LensB}
}
%--------------------------------------------------%
where $\hE^{\rm WF}_{\l m}$ and $\estk^{\rm WF}_{LM}$ are the Wiener-filtered observed $E$-mode polarization and lensing convergence, respectively. In idealistic cases, we compute the Wiener-filtered multipoles as
%--------------------------------------------------%
\al{
    \hE^{\rm WF}_{\l m} = \frac{C_\l^{EE}}{C^{EE,{\rm obs}}_\l}\hE_{\l m} 
    \,, \\
    \estk^{\rm WF}_{LM} = \frac{C_L^{\kappa m}}{C^{mm,{\rm obs}}_L}\widehat{m}_{LM} 
    \,, 
}
%--------------------------------------------------%
where $m$ is a lensing mass tracer. The $\hE$ and $\widehat{m}$ denote the observed $E$-mode polarization and $m$, respectively, $C^{EE}_\l$ is the $E$-mode power spectrum, $C_L^{\kappa m}$ is the cross-power spectrum between $\kappa$ and $m$, and $C_l^{EE,{\rm obs}}$ and $C^{mm,{\rm obs}}_L$ are the best estimates of the angular power spectra of $\hE$ and $\widehat{m}$, respectively. The template provides a good approximation for the large-scale $B$-mode polarization \cite{Challinor:2005:lensfull,BaleatoLizancos:2020:template-method}. Using this template, we can remove part of the lensing $B$-mode polarization from the observed map \cite{Seljak:2003:delens}, or construct the multi-spectrum likelihood \cite{BKSPT,Namikawa:2021:SO-delens} to improve constraints on $r$.

\subsection{Lensing mass tracers}

To estimate the lensing mass map, we can use the internally reconstructed lensing map from \LB\ observation. However, the \LB\ lensing mass map alone cannot remove the lensing $B$-mode polarization efficiently \cite{Diego-Palazuelos:2020:delens}. A more optimal way is to employ the lensing convergence field estimated from observations of the large-scale structure tracers, such as the spatial distribution of galaxies or the CIB~\cite{Sherwin:2015,Simard:2015,Namikawa:2015:delens,Karkare:2019:delens}. The delensing efficiency of these tracers depends on the correlation coefficients between the CMB lensing signal and the observed mass tracer. In this section, we provide our formulas for computing the correlation coefficients and assumptions for the lensing mass tracers. 

\subsubsection{Angular power spectrum}

We compute the angular auto and cross-power spectra of the CMB lensing, CIB, and galaxies as follows. In a spatially flat cosmological model, the angular power spectra are given by (see, e.g., Refs.~\cite{Hu:2001:darksynergy,Lewis:1999:camb})
%----------%
\al{
	C^{XY}_L &= 4\pi \INT{}{\ln k}{}{0}{\infty} 
		\INT{}{\chi}{}{0}{\infty} j_L(k\chi) \INT{}{\chi'}{}{0}{\infty} j_L(k\chi') 
		\notag \\
	&\quad \times W^X(k,\chi) W^Y(k,\chi') \Delta_{\rm m}(k;\chi,\chi')
	\,, \label{Eq:mass:aps}
}
%----------%
with the quantity $\chi$ being the comoving radial distance. Here, $X$ and $Y$ denote the observables from either the CMB lensing ($\kappa$), galaxy number density fluctuations ($g$), or CIB ($I$). The function $\Delta_{\rm m}(k;\chi,\chi')$ is the dimensionless power spectrum of the matter density fluctuations, and $j_\l$ is the spherical Bessel function. The function $W^{\rm X}(k,\chi)$ is the weight function of an observable X, the functional form of which will be specified for each mass tracer as follows. 

\subsubsection{CMB Lensing}

Since the lensing comes from the last-scattering surface of the CMB photons which is approximately described by a single-source plane, the weight function of the lensing convergence is given by (e.g., Ref.~\cite{Lewis:2006:review})
\al{
	W^{\kappa}(k,\chi) = -\frac{L(L+1)}{2}\frac{3\Omega_{\rm m,0}H_0^2(1+z)}{k^2}\frac{\chi_*-\chi}{\chi_*\chi} \qquad (0\leq \chi\leq\chi_*) 
	\,, 
}
and becomes zero otherwise. Here, the quantity $\chi_*$ denotes the comoving radial distance to the last-scattering surface, $\Omega_{\rm m,0}$ is the fractional energy density of matter at present, and $H_0$ is the current expansion rate.

\subsubsection{Galaxy number density} \label{Sec:redshift-bins}

We follow the theoretical modeling of the galaxy number density fluctuations described in, e.g., Refs.~\cite{Yu:2017:delens,Namikawa:2021:SO-delens}. The weight function of the galaxy number density fluctuations in the $i$th redshift bin becomes 
%----------%
\al{
	W^{{\rm g},i}(\chi) = \D{n^i_{\rm gal}}{\chi}(z(\chi)) \,b_{\rm gal}(z(\chi)) 
	\,, 
}
%----------%
where ${\rm d}n^i_{\rm gal}/{\rm d}\chi$ and $b_{\rm gal}(z)$ are the normalized redshift distribution of galaxies in the $i$th bin and the linear galaxy bias, respectively. For the redshift distribution function, we adopt the following form \cite{Euclid12,LSST}: 
%----------%
\al{
	\D{n^i_{\rm gal}}{\chi}(z) =
    \frac{\beta H(z)}{\Gamma[(\alpha+1)/\beta]}\frac{z^\alpha}{z_0^{\alpha+1}}\exp\left[-\left(\frac{z}{z_0}\right)^\beta\right] p^i_{\rm gal}(z,\sigma_z)
	\,, \label{Eq:ngal}
}
where $\alpha$, $\beta$, and $z_0$ are parameters determining the distribution. The parameter $z_0$ is related to the mean redshift $z_{\rm m}$ as 
\al{
    z_{\rm m} = \frac{\Gamma[(\alpha+2)/\beta]}{\Gamma[(\alpha+1)/\beta]}z_0
    \,. 
}
%----------%
$H(z)$ is the expansion rate, simply from the conversion between $z$ and $\chi$. The function, $p^i_{\rm gal}(z,\sigma_z)$, specifies the redshift distribution of galaxies in $i$th redshift bin with the modification of the redshift distribution by the photometric redshift errors. We assume that $p^i_{\rm gal}(z,\sigma_z)$ has the following form: 
\al{
    p^i_{\rm gal}(z,\sigma_z) = \frac{1}{2}\left[{\rm erfc}\left(\frac{z_{i-1}-z}{\sqrt{2}\sigma(z)}\right)-{\rm erfc}
	\left(\frac{z_i-z}{\sqrt{2}\sigma(z)}\right)\right] 
	\,, 
}
where $\sigma(z)=\sigma_z(1+z)$, $z_{i-1}$ and $z_i$ are the minimum and maximum photometric redshift of the $i$th bin, and the function ${\rm erfc}(x)$ is the complementary error function defined by
\al{
	{\rm erfc}(x)\equiv \frac{2}{\sqrt{\pi}}\INT{}{z}{}{x}{\infty} \E^{-z^2}
	\,.
}
In this paper, we consider \Euclid\ and LSST as galaxy surveys. For \Euclid, we consider $\alpha=2$, $\beta=1.5$, $\sigma_z=0.05$, $z_0=0.9/\sqrt{2}$, and $b_{\rm gal}(z)=\sqrt{1+z}$ \cite{Euclid12} and the following five tomographic bins; $[0,0.8,1.5,2.0,2.5,6.]$. 
For the LSST, we follow Refs.~\cite{LSST,Yu:2017:delens} and choose $\alpha=2$, $\beta=1$, $z_0=$, $\sigma_z=0.05$, $b_{\rm gal}(z)=1+0.84z$, and the five redshift tomographic bins:  $[0,0.5,1.0,2.0,3.0,6.]$.

\subsubsection{Cosmic Infrared Background}

For the CIB, we assume a model to mimic the result of Ref.~\cite{Yu:2017:delens}, where they consider the CIB map constructed via the Generalized Needlet Internal Linear Combination (GNILC) algorithm applied to the \Planck\ PR2 data \cite{Remazeilles:2011:NILC,P16:GNILC}. For a theoretical computation of the CIB angular power spectrum, they adopt the single spectral energy distribution (SED) model suggested by the \Planck\ observations \cite{PR2:phi,PR1:CIBxphi,PR1:CIB}. Specifically, the window function is defined as 
%----------%
\al{
	W^I(\chi) = b_{\rm c}\frac{\chi^2}{(1+z)^2}f[\nu(1+z)]\exp\left[-\frac{(z-z_{\rm c})^2}{2\sigma_z^2}\right] 
	\,, 
}
%----------%
where $z_{\rm c}=\sigma_z=2$, $b_{\rm c}$ is a normalization factor, $z$ is the redshift as a function of $\chi$, and
%----------%
\al{
    f(\nu) = \begin{cases}
        (e^{h\nu/k_{\rm B}T}-1)^{-1}\nu^{\beta+3} & (\nu\leq\nu')
    \,, \\        
        (e^{h\nu'/k_{\rm B}T}-1)^{-1}\nu'^{\beta+3}\left(\frac{\nu}{\nu'}\right)^{-\alpha} & (\nu>\nu')
    \,, 
    \end{cases}
}
%----------%
with $T=34$\,K, $\alpha=\beta=2$, $\nu=353$\,GHz and $\nu'=4955$\,GHz.

\subsubsection{Kernel function} \label{Sec:kernel}

\begin{figure}[t]
\centering
\includegraphics[width=100mm]{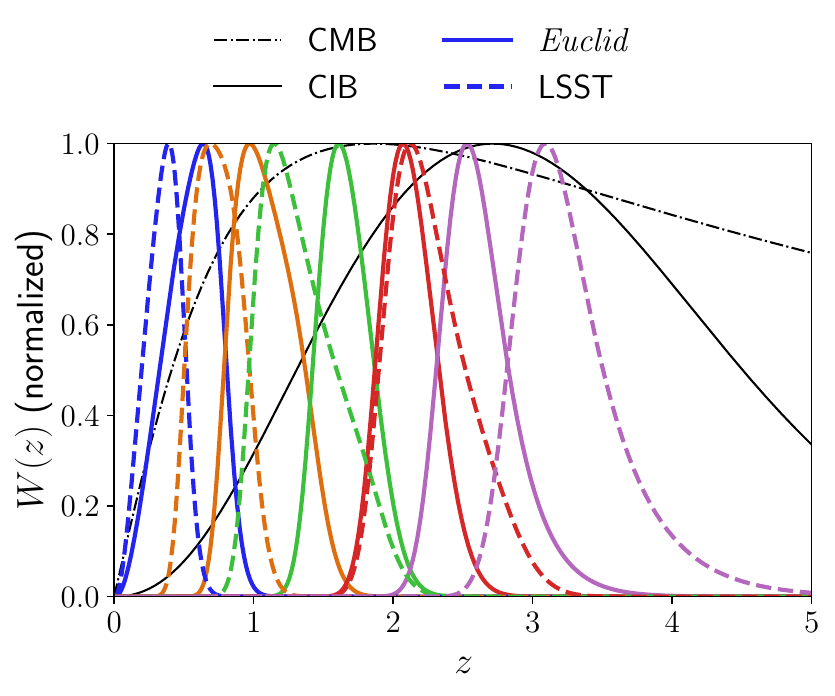}
\caption{
Kernel function for CMB lensing, CIB, \Euclid, and LSST galaxies (see text in Sec.~\ref{Sec:kernel} for the definitions). The different colors represent different redshift bins for galaxy surveys defined in Sec.~\ref{Sec:redshift-bins}. The function is normalized so that the peak is unity. 
}
\label{fig:kernel}
\end{figure}

Figure \ref{fig:kernel} shows the kernel functions for each mass tracer defined as follows. For CMB lensing, we show $-(k\chi)^2W^\kappa/(L(L+1))/H(z)$. For CIB and galaxies, we show $W^I/H(z)$ and $W^{g,i}/H(z)$, respectively. The galaxies provide a lensing mass map at lower redshifts, while the CIB data have information on relatively higher redshifts. Thus, combining these tracers will provide a lensing mass map for a wide redshift range.

\subsection{Correlation coefficients}

\begin{figure}[t]
\centering
\includegraphics[width=75mm]{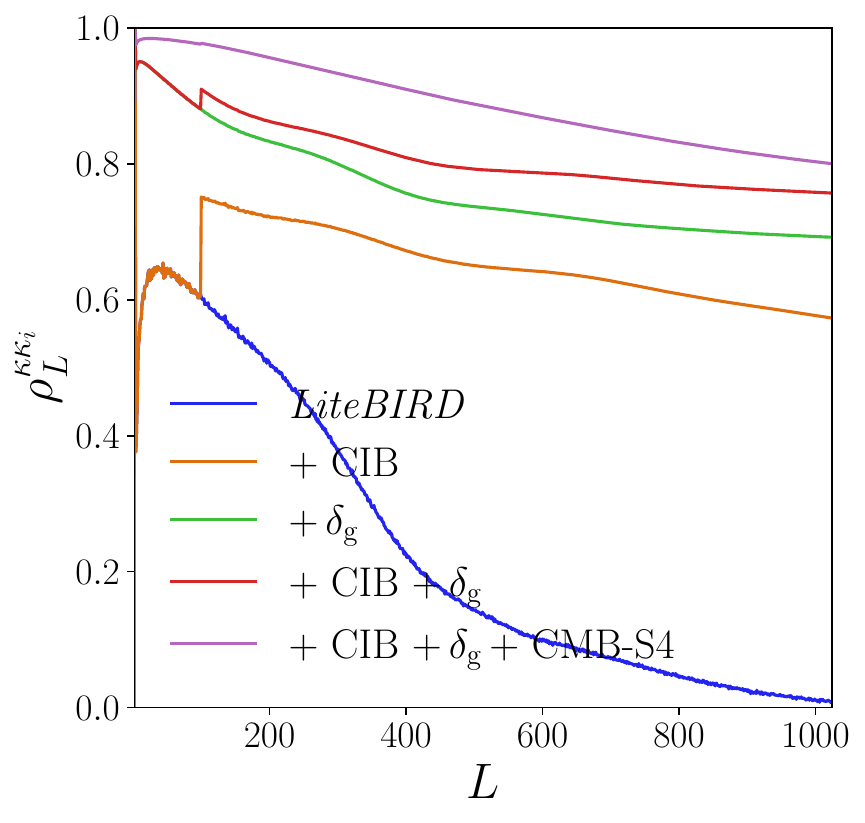}
\includegraphics[width=75mm]{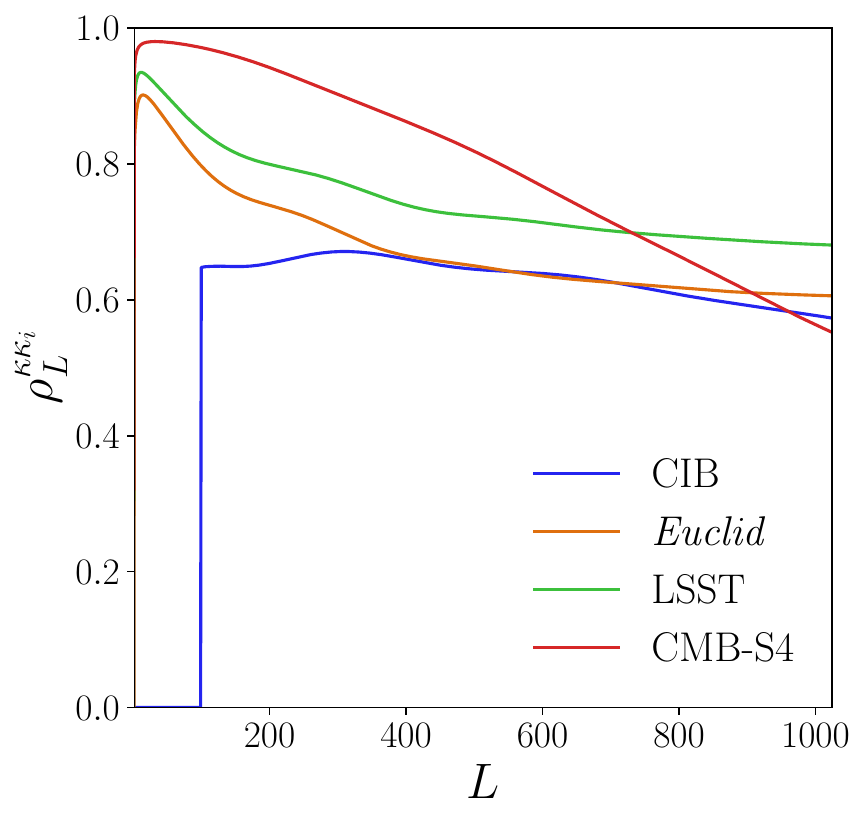}
\caption{
{\it Left}: Correlation coefficients between the ideal full-sky CMB lensing and observed lensing mass-tracer maps obtained by combining several of the \LB\ lensing maps with the CIB, galaxy number density from \Euclid\ and LSST, and \S4\ lensing maps. 
{\it Right}: Same as the left panel but for the correlation coefficients for each external data set taken separately, i.e., the CIB, galaxies from \Euclid\ and LSST, and \S4.
}
\label{fig:rho}
\end{figure}

From the theoretical angular power spectra, we can estimate how significantly each mass tracer correlates with the true CMB lensing mass map. To assess the significance of the correlation, we introduce the correlation coefficients of an observed mass tracer map with the true CMB lensing mass map:
\al{
    \rho^{\kappa\kappa^i}_L \equiv \frac{C_L^{\kappa\kappa^i}}{\sqrt{C_L^{\kappa\kappa}C_L^{\kappa^i\kappa^i,{\rm obs}}}}
    \,.
}
Here, $C_L^{\kappa\kappa^i}$ is the cross-power spectrum between CMB lensing and $i$th observed lensing mass maps, $C_L^{\kappa\kappa}$ is the CMB lensing power spectrum, and $C_L^{\kappa^i\kappa^i,{\rm obs}}$ is the auto-power spectrum of the $i$th observed lensing mass map, which contains both signal and noise. 
We compute the signal angular power spectra defined in Eq.~\eqref{Eq:mass:aps} with {\tt CAMB} \cite{Lewis:1999:camb} by providing the appropriate window functions given above. To compute the correlation coefficients for the combined cases, we combine the mass tracers following Ref.~\cite{Sherwin:2015}. For the CIB, we employ the same noise spectrum given by Ref.~\cite{Yu:2017:delens} and remove large-scale CIB data ($L<100$) to avoid contamination from Galactic foreground residuals. For galaxy surveys, we employ a shot noise determined by the number density of galaxies and assume the total galaxy number density of $30$ arcmin$^{-2}$ \cite{Euclid12} and $40$ arcmin$^{-2}$ \cite{LSST} for \Euclid\ and LSST, respectively. Note that we also check cases in which we increase the number of the tomographic bins, but the correlation coefficients are almost unchanged. 

Figure \ref{fig:rho} shows the results of the correlation coefficients. For the large-scale $B$-mode delensing, the most important scale for efficient delensing is $L\simeq 300$ -- $500$ \cite{Simard:2015,Sherwin:2015}. From the left panel, adding external mass tracers significantly improves the correlation coefficient and the delensing efficiency of the large-scale $B$-mode polarization. In the right panel, we show that the correlation coefficients of CIB and \Euclid\ galaxies are not better than that of the LSST galaxies at $L\simeq 400$. However, the sky coverage of LSST is limited compared to other mass tracers, and a net contribution to the improvement on $r$ constraints with the CIB becomes not so different from that obtained with galaxies.

\section{Simulation setup} \label{sec:simulation}

\begin{figure}[t]
\centering
\includegraphics[width=70mm]{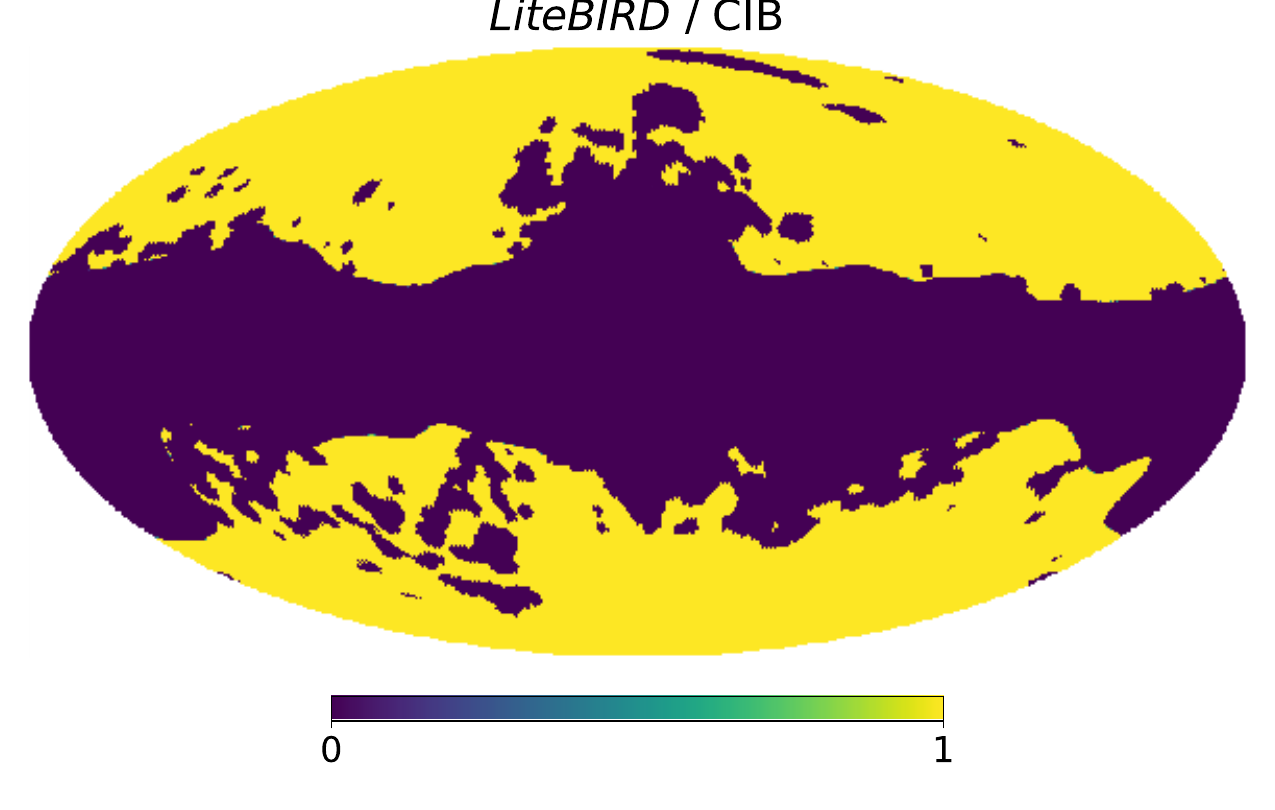}
\includegraphics[width=70mm]{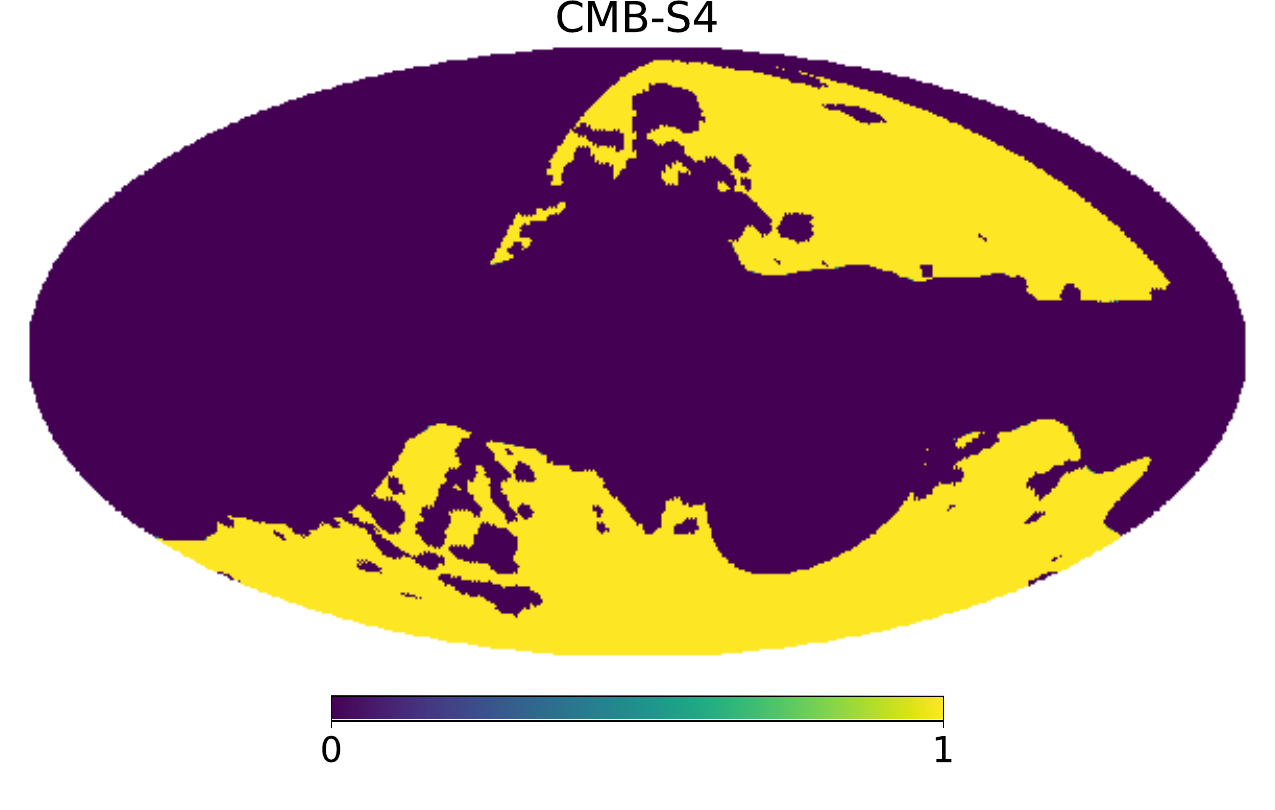}
\includegraphics[width=70mm]{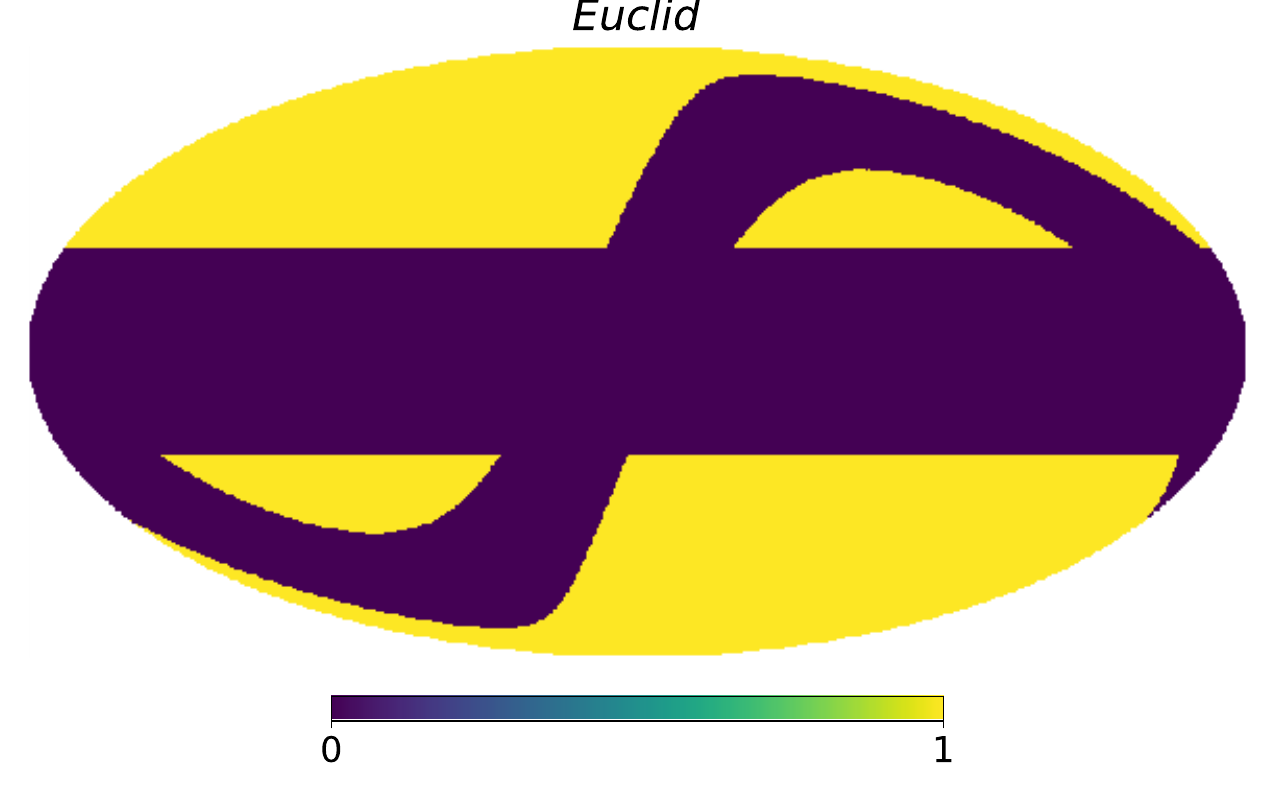}
\includegraphics[width=70mm]{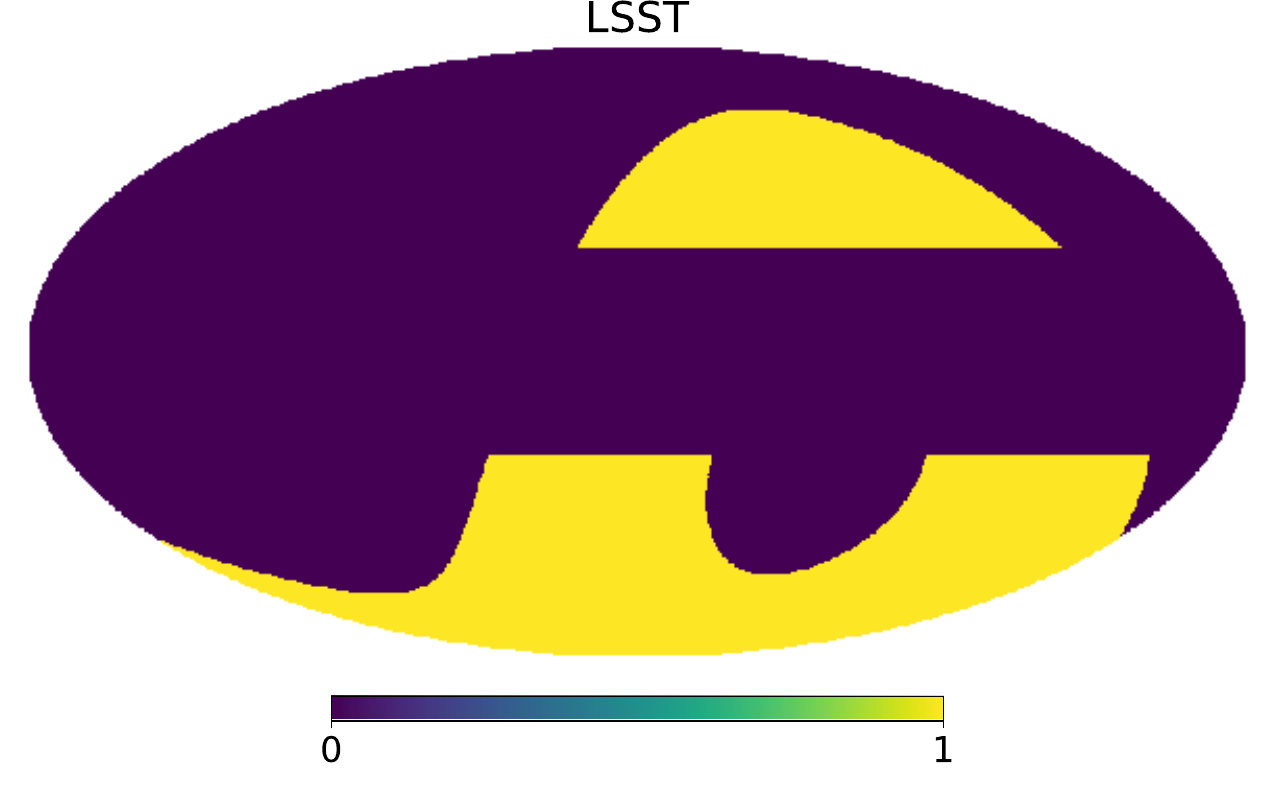}
\caption{
Survey windows for each observation in Galactic coordinate used in our analysis. 
}
\label{fig:window}
\end{figure}

\subsection{Survey window}

We plot the survey windows in Fig.~\ref{fig:window}. 
For the \LB\ data, we use the same Galactic mask used in \pteppaper\ for the \LB\ observation, which covers approximately $50\%$ of the sky. We also apply the same mask to the CIB map to mimic the sky fraction of the survey window used in Ref.~\cite{Yu:2017:delens}. 
For the \S4\ observation, we create a survey window that mimics the \S4\ LAT survey window multiplied by the \LB\ Galactic mask to avoid the Galactic plane. In practice, we would need further masks to avoid extragalactic sources, but we ignore these masks in this paper. 
For \Euclid\ and LSST, we follow Refs.~\cite{Euclid:mask} and \cite{LSST}, respectively, to create the survey window. 
We further multiply the \LB\ Galactic mask to those of the \Euclid\ and LSST survey windows to consider only the region overlapping with the galaxy number density map. 

\subsection{Wiener-filtered {\it E}-modes}

To obtain the Wiener-filtered $E$-mode polarization from the component-separated polarization map, we use the same method described in \paperI\ for the \LB-only case. We also consider the case of combining \LB\ and \S4\ data for delensing. In this case, the Wiener-filtered $E$-mode polarization maps are obtained by solving the following equation~\cite{Eriksen:2004:wiener}: 
\al{
    & \left(1+\sum_{t}\bR{C}^{1/2}\bR{B}_t\bR{Y}_2^\dagger\bR{N}^{-1}_t\bR{Y}_2\bR{B}_t\bR{C}^{1/2}\right) (\bR{C}^{-1/2}\bm{\hpol}^{\rm WF}) 
	= \sum_{t}\bR{C}^{1/2}\bR{B}_t\bR{Y}^\dagger_2 \bR{N}^{-1}_t \bm{\hpol}^{\rm obs}_t
	\,. \label{Eq:Cinv}
}
Here, $t$ corresponds to indices for the input maps specifying the experiment (\LB\ or \S4), the vector $\bm{\hpol}^{\rm WF}$ has the harmonic coefficients of the Wiener-filtered $E$- and $B$-mode polarization, $\bR{C}$ is the diagonal signal covariance of the lensed $E$- and $B$-mode polarization in spherical-harmonic space, and $\bR{B}_t$ is a diagonal matrix to include the beam smearing. The matrix $\bR{C}^{1/2}$ is defined so that its square is equal to $\bR{C}$. The real-space vector $\bm{\hpol}^{\rm obs}_t$ contains the Stokes $Q$ and $U$ maps observed by experiment $t$, and $\bR{N}_t$ is the pixel-space covariance matrix of the instrumental noise in these maps. The matrix $\bR{Y}_2$ is defined to transform the multipoles of the $E$- and $B$-mode polarization into real-space maps of the Stokes parameters $Q$ and $U$. Note that this filtering optimally mitigates the $E$-to-$B$ leakage due to the masking. We use the code implemented in {\tt cmblensplus} \cite{cmblensplus}. 

To generate an observed CMB map from a \S4-like experiment, we use the same realizations of the input lensed CMB polarization given in \paperI\ by simply adding a random Gaussian noise component generated by the polarization noise spectrum obtained from the \S4\ wiki.\footnote{\url{https://cmb-s4.uchicago.edu/wiki/index.php/Survey_Performance_Expectations}}
The noise curve contains the effect of the foreground cleaning and atmospheric $1/f$ noise. We assume the standard internal-linear-combination noise spectrum for this paper. The polarization map is then multiplied by the \S4\ mask described above and combined with the \LB\ polarization map to obtain the optimal $E$-mode polarization. 
We assign infinite noise in the noise covariance for unobserved pixels.
In this paper, we do not include any extragalactic foregrounds, but in practice, we should include masks for extragalactic contaminants as unobserved pixels, which is left for future work.

\subsection{Combining mass tracers}

Next, we discuss an estimate of lensing mass for \LB\ delensing analysis in the presence of different sky coverage of mass tracers. Let us assume that we have $n$ observed mass tracers, $\kappa^{{\rm obs},1},\kappa^{{\rm obs},2},\dots,\kappa^{{\rm obs},n}$. We obtained the combined lensing-mass map as
\al{
    \estk^{\rm WF}_{LM} = \sum_{i=1}^n C^{\kappa\kappa^i}_L\estk^{\rm inv,i}_{LM}
    \,, 
}
where we obtain the inverse-variance filtered lensing-mass tracers, $\estk^{\rm inv,i}_{LM}$, as follows: 
\al{
	&\left(1+\bR{Y}^\dagger_0\ol{\bR{N}}^{-1}_\kappa\bR{Y}_0\bR{C}\right) \bm{\estk}^{\rm inv} 
	= \bR{Y}^\dagger_0\ol{\bR{N}}^{-1}_\kappa\bm{\estk}^{\rm obs}
	\,. \label{Eq:Cinv:kappa}
}
Here, $\bm{\estk}^{\rm obs}$ is the data vector of the lensing mass maps in pixel space. The lensing-mass signal covariance is block diagonal in $(L,M)$, and each block matrix is symmetric: 
\al{
    \{\bR{C}\}_{LM,L'M'} \equiv \delta_{LL'}\delta_{MM'}\begin{pmatrix} 
    C^{\kappa^1\kappa^1}_L & \cdots & C^{\kappa^1\kappa^n}_L \\ 
    \vdots & & \vdots \\ 
    C^{\kappa^1\kappa^n}_L & \cdots & C^{\kappa^n\kappa^n}_L 
    \end{pmatrix} 
    \,. \label{Eq:kappa:cov}
}
The matrix $\bR{Y}_0$ converts the multipoles of the lensing convergence to the lensing mass map. The matrix $\ol{\bR{N}}_\kappa$ denotes the pixel-pixel noise covariance. 

We assume that the pixel-pixel noise covariance matrix for each lensing mass map is given by
\al{
    \{\ol{\bR{N}}_\kappa\}^i = \bR{M}_i\bR{Y}_0\bR{N}^i_\kappa\bR{Y}^\dagger_0\bR{M}_i
    \,, 
}
where $\bR{M}_i$ is a diagonal matrix for the survey binary mask of the $i$th mass tracer, and the noise covariance in harmonic space, $\bR{N}^i_\kappa$, is defined as a diagonal matrix: 
\al{
    \{\bR{N}^i_\kappa\}_{LM,LM} = N^i_L 
    \,. 
}
Here, $N^i_L$ is the noise power spectrum of the $i$th lensing mass map. For \LB\ and \S4, $N^i_L$ is the disconnected bias for the reconstructed CMB lensing maps, which we approximate by the estimator normalization. Although the noise of CMB lensing maps from \LB\ and \S4\ could be correlated, we ignore such correlations because the lensing signals of \LB\ and \S4\ are mostly reconstructed from CMB anisotropies at large ($\l\alt 500$) and small angular scales ($\l\agt 500$), respectively, and the correlation of the two reconstruction noise terms would be negligible. In the idealistic full-sky observation where $\bR{M}_i=1$, the Wiener-filtered lensing mass map becomes that obtained by Ref.~\cite{Sherwin:2015}. We use the code implemented in {\tt cmblensplus} \cite{cmblensplus}. 

We use the \LB\ reconstructed lensing mass map of \paperI\ but exclude $B$-mode polarization at multipoles $\l\leq 190$ for the lensing reconstruction to avoid the delensing bias, which primarily originated from the fact that $B$-mode polarization used for the lensing reconstruction correlates with that to be delensed (see, e.g., Refs.~\cite{Namikawa:2017:delens,BaleatoLizancos:2020:delensbias} for more detailed discussion). The details of the simulation production and experimental setups for \LB\ for the lensing reconstruction are given in \paperI. 
We also generate the CIB and galaxy tracers in the full sky as a constrained realization of multiple random Gaussian fields for a given realization of the input CMB lensing map. We do not consider the nonlinear evolution of the external tracers since its impact would be small \cite{Namikawa:2018:nldelens}.

\subsection{Large-scale LiteBIRD {\it B}-modes}

For the delensing of the large-scale $B$-modes, we prepare the $Q$ and $U$ polarization maps, which contain lensed CMB and Galactic foreground noise. We optionally further include the IGW contribution to the polarization map. The input full-sky lensed CMB of \paperI\ are projected onto a map with $N_{\rm side}=128$ \cite{Gorski:2004:healpix} resolution, including the pixel-window convolution and $80$-arcmin Gaussian beam convolution. 
%Note that the Gaussian beam convolution matches the Galactic foreground noises provided by \pteppaper\ described below. 
The IGW $B$-mode polarization is also generated as a random Gaussian field with $r=1$ and projected onto a map with the same map projection. We use the noise realizations after the Galactic foreground cleaning obtained from \pteppaper. The maps are provided at $N_{\rm side}=64$ resolution with $80$-arcmin Gaussian beam convolution, and we transform the resolution to $N_{\rm side}=128$ to match the lensed CMB maps. 

Note that the noise realizations used for the lensing reconstruction and lensing $B$-mode template construction are independent of \pteppaper. This means that we ignore the correlations between the noise involved in the lensing $B$-mode template and that in the large-scale $B$-modes to be delensed. The correlation is negligible as long as we use different CMB multipoles for the lensing $B$-mode template and $B$-mode polarization to be delensed, as demonstrated by multiple existing works \cite{Seljak:2003:delens,Namikawa:2017:delens,Beck:2020:FG-lens}.

\section{Results} \label{sec:results}

\subsection{Delensed {\it B}-mode power spectrum}

\begin{figure}[t]
\centering
\includegraphics[width=75mm]{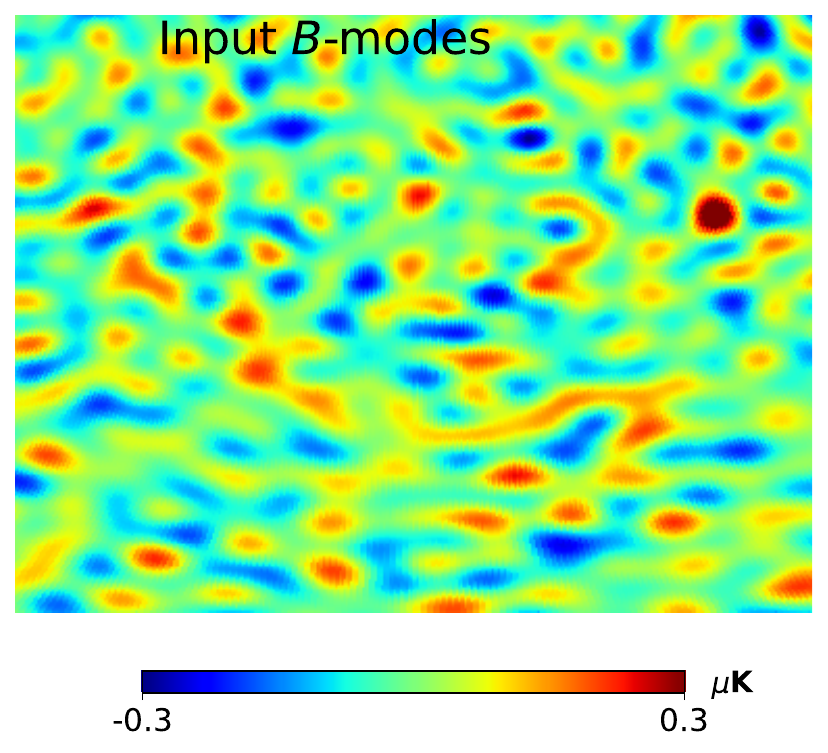}
\includegraphics[width=75mm]{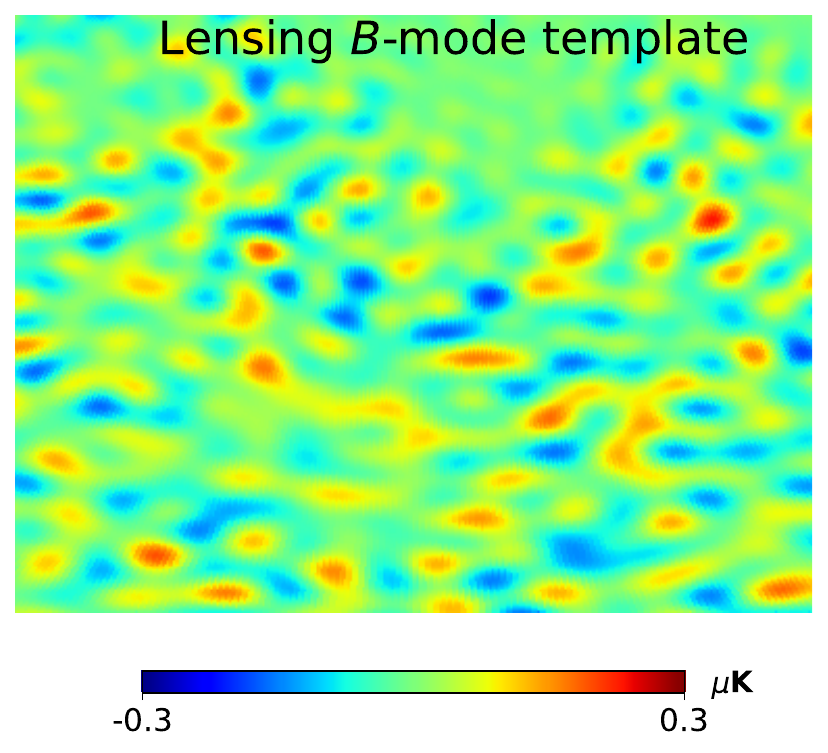}
\caption{
Partial maps of the full-sky $B$-mode polarization (Left) and lensing $B$-mode template (Right). For the lensing $B$-mode template, we use mass tracers from the \S4\ lensing map, the CIB, galaxies, and the internal lensing map. 
}
\label{fig:bmap}
\end{figure}

We first show the lensing $B$-mode template in Fig.~\ref{fig:bmap}. We use mass tracers from the \S4\ lensing map, the CIB, galaxies, and the internal lensing map. We compute the input $B$-mode map by transforming the input $B$-mode harmonic coefficients using the spin-$0$ inverse spherical harmonics. We include multipoles $10\leq\l\leq 190$ for visualization. We can see a clear correlation between the input $B$-mode map and lensing $B$-mode template in Fig.~\ref{fig:bmap}.

\begin{figure}[t]
\centering
\includegraphics[width=100mm]{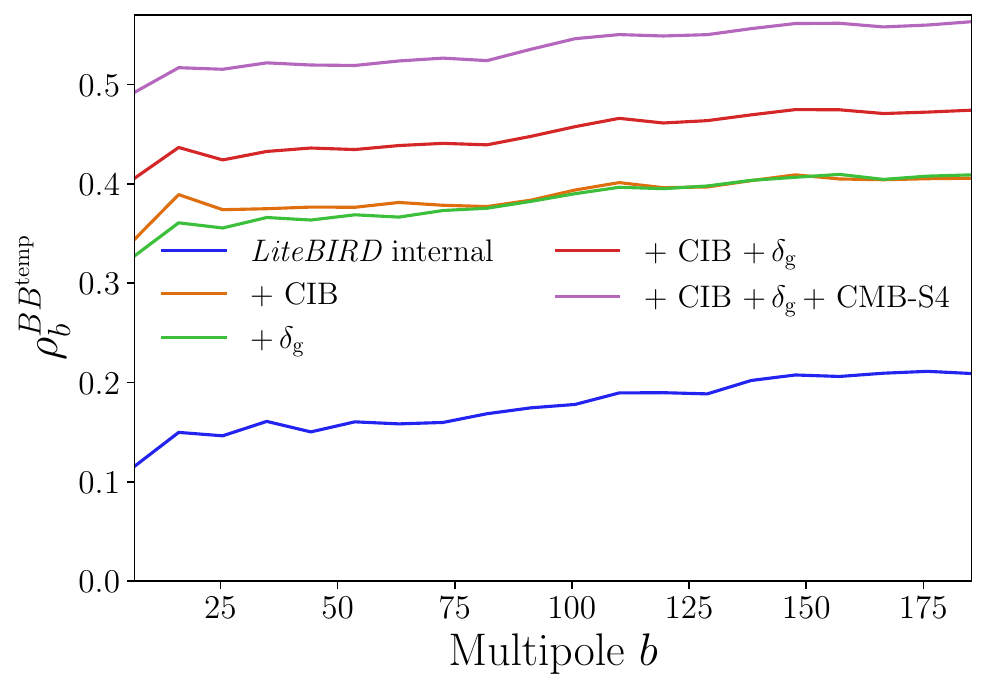}
\caption{
Correlation coefficients between the input full-sky $B$-mode polarization and the lensing $B$-mode template. We show the cases using the mass tracers of the \LB\ lensing map (``\LB\ internal"), adding CIB (``$+$CIB"), adding galaxy number density fluctuations (``$+\delta_{\rm g}$"), adding CIB and galaxy number density fluctuations (``$+$CIB$+\delta_{\rm g}$"), and further adding \S4\ $E$-mode polarization and lensing map (``$+$CIB$+\delta_{\rm g}+$CMB-S4"). 
}
\label{fig:rhobb}
\end{figure}

\begin{figure}[t]
\centering
\includegraphics[width=75mm]{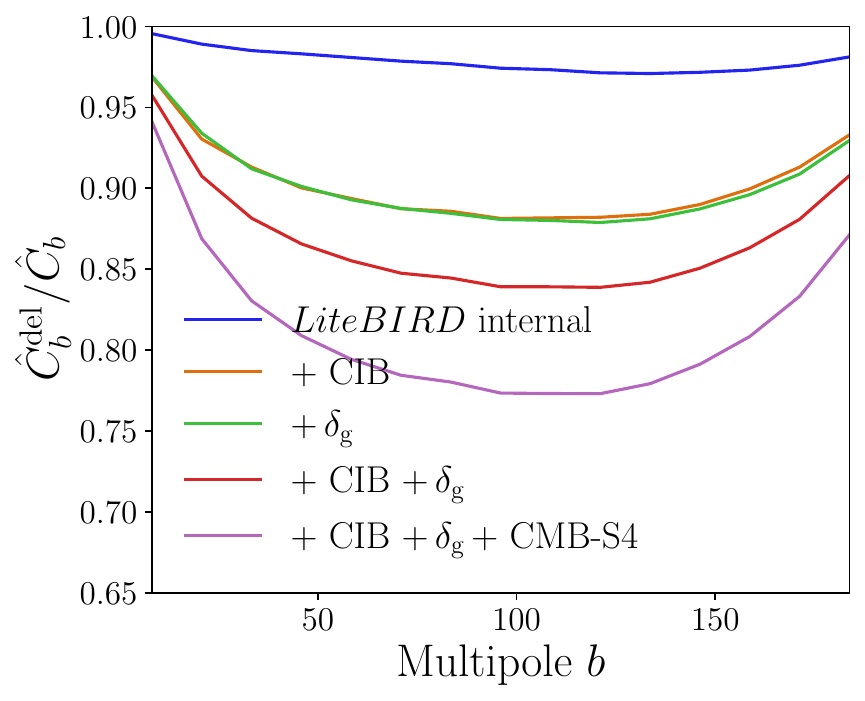}
\includegraphics[width=75mm]{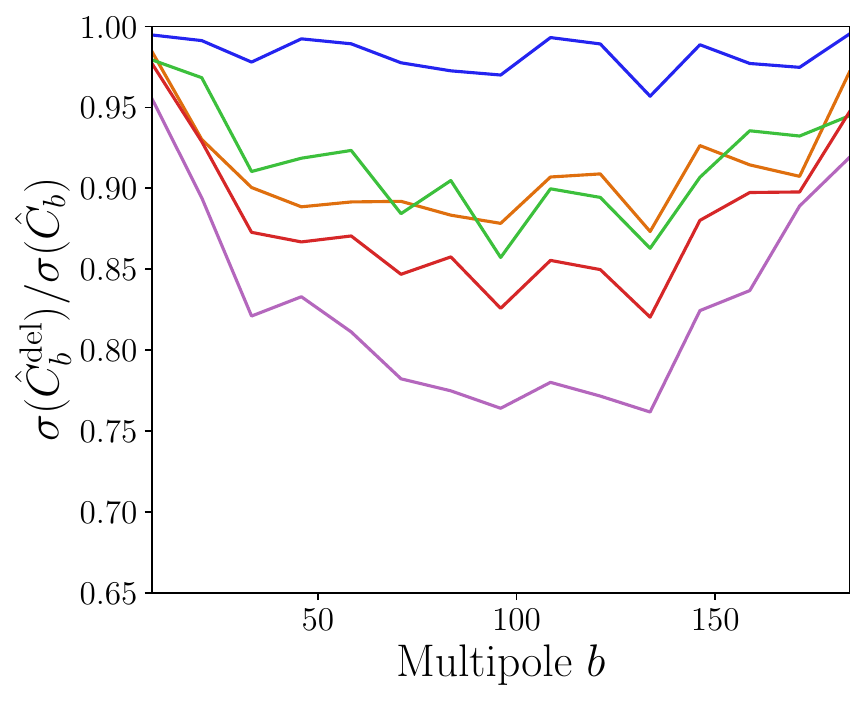}
\caption{
{\it Left}: Same as Fig.~\ref{fig:rhobb}, but for the ratio of the delensed $B$-mode power spectrum to the non-delensed $B$-mode power spectrum. 
{\it Right}: Same as {\it Left}, but for the $1\,\sigma$ error at each multipole bin. 
}
\label{fig:clbbvlbb}
\end{figure}

We also show the efficiency of delensing by computing the delensed $B$-mode power spectrum. For a given realization of the \LB\ large-scale $B$-mode polarization, we compute the Wiener-filtered $B$-mode polarization, $\hat{B}^{\rm WF}_{l m}$, as described in Eq.~\eqref{Eq:Cinv}, to reconstruct the $B$-modes optimally and to mitigate the $E$-to-$B$ leakage induced by the survey window. We then compute the angular power spectrum of the Wiener-filtered $B$-mode polarization as $\hat{C}_l=\sum_m |\hat{B}^{\rm WF}_{l m}|^2/(2l+1)$ and obtain the binned power spectrum, $\hat{C}_b$. For multipole binning, we divide the multipoles between $\l=2$ and $190$ into $15$ bins. 

For the delensing case, we define the delensed $B$ modes as
\al{
    B^{\rm del}_{\l m} = \hat{B}_{\ell m} - \alpha_\l B^{\rm temp}_{\l m}
    \,, 
}
where $\alpha_\l$ is determined so that the variance of $B^{\rm del}$ is minimized, i.e., $\alpha_\l=C^{BB,{\rm cross}}_\l/C_\l^{BB,{\rm temp}}$. Here, $\hat{C}_\l^{\rm cross}$ is the cross-power spectrum between the Wiener-filtered $B$-mode polarization and lensing $B$-mode template, and $\hat{C}_\l^{\rm temp}$ is the lensing template power spectrum. We thus compute the binned angular power spectrum of the delensed $B$-mode power spectrum as
\al{
    \hat{C}^{\rm del}_b = \hat{C}_b - 2\alpha_b\hat{C}^{\rm cross}_b + \alpha_b^2 \hat{C}^{\rm temp}_b 
    \,,
}
where $\alpha_b=\ave{\hat{C}^{\rm cross}_b}/\ave{\hat{C}^{\rm temp}_b}$ is evaluated from simulations. 

Figure \ref{fig:rhobb} shows the correlation coefficients between the input full-sky $B$-mode polarization and lensing $B$-mode template constructed from mass tracers. We show the cases using the \LB\ lensing map, adding CIB, adding galaxy surveys, adding both CIB and galaxy surveys, and further adding the \S4\ $E$-mode polarization and lensing map. The correlation coefficient for the \LB\ case is much lower than other cases. Combining all available mass tracers will provide a lensing $B$-mode template that correlates with the full-sky input $B$-mode map by roughly $50\%$ at $\l<100$. 

Figure \ref{fig:clbbvlbb} shows the $B$-mode power spectrum ratio with delensing compared to that without delensing. The \LB\ lensing map does not help much in removing the lensing $B$-mode power spectrum. The lensing $B$-mode power is reduced by $15\%$ at $50\alt\l\alt 150$ if we use both CIB and galaxies and is further reduced by $20\%$ in the same multipole range if we combine with \S4\ data. 
Figure \ref{fig:clbbvlbb} also compares the $1\,\sigma$ error of the $B$-mode power spectrum with delensing to the one without delensing. 
The ratio has a similar trend to the one shown in the Left panel of the same figure.

\subsection{Improvement on constraining tensor-to-scalar ratio}

\begin{figure}[t]
\centering
\includegraphics[width=100mm]{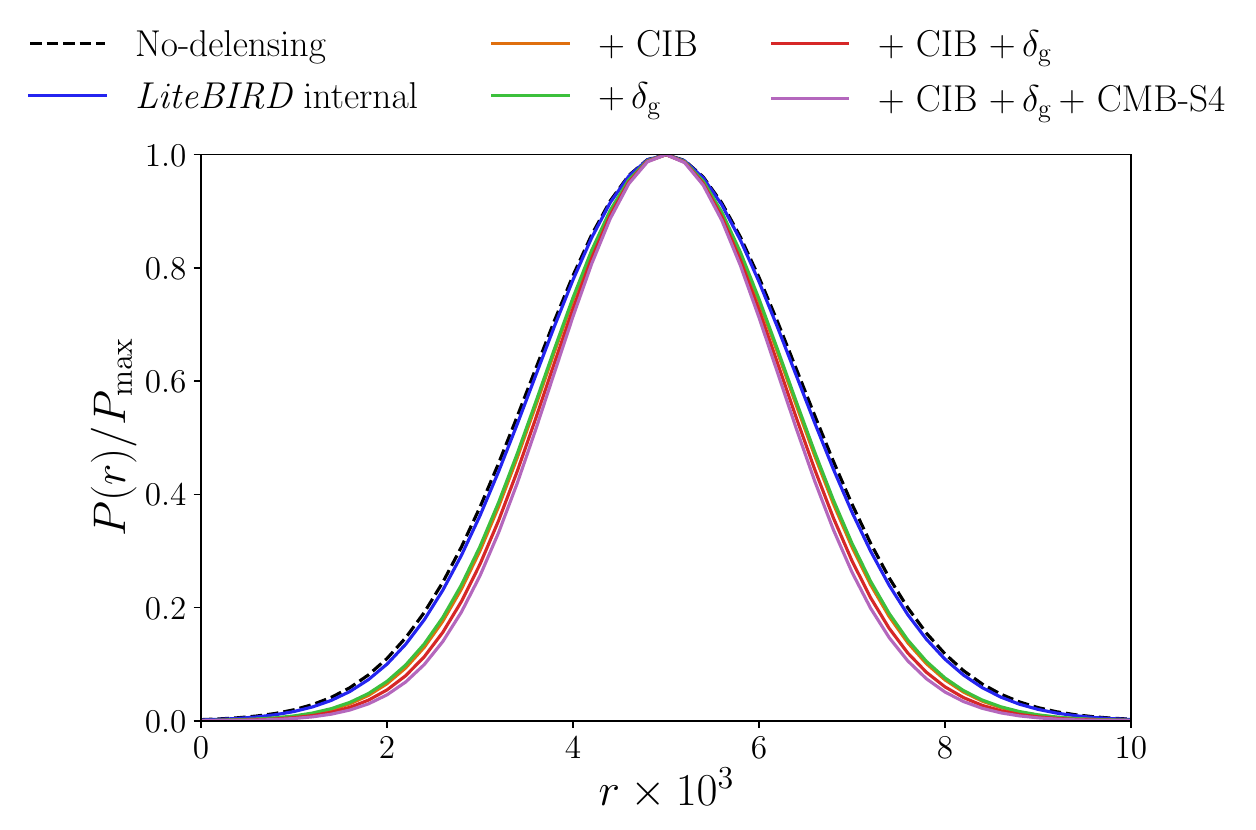}
\caption{
Posterior distribution of $r$ normalized to unity \tn{where the fiducial value is $r=5\times 10^{-3}$}. We show the cases without (No-delensing) and with delensing using \LB\ data alone (\LB\ internal), adding the CIB (+CIB), galaxy surveys (+$\delta_{\rm g}$), the CIB and galaxy surveys (+CIB+$\delta_{\rm g}$), and combining \LB, \S4, the CIB and galaxies (+CIB+$\delta_{\rm g}$+\S4). 
}
\label{fig:posterior}
\end{figure}

\begin{table}[t]
 \centering
 \begin{tabular}{lc}
 \hline 
 & $\sigma(r) \times 10^3$ \\ \hline
 No-delensing  & 1.44 \\ \hline
 \LB\ internal & 1.41 \\ 
 + CIB & 1.30 \\ 
 + $\delta_{\rm g}$ & 1.31 \\ 
 + CIB + $\delta_{\rm g}$ & 1.25 \\
 + CIB + $\delta_{\rm g}$ + \S4\ & 1.21
 \end{tabular}
 \caption{Expected constraints on $r$ without (No-delensing) and with delensing using \LB\ data alone (\LB\ internal), adding the CIB (+CIB), galaxy surveys (+$\delta_{\rm g}$), the CIB and galaxy surveys (+CIB+$\delta_{\rm g}$), and combining \LB, \S4, the CIB and galaxies (+CIB+$\delta_{\rm g}$+\S4). \tn{We show the $1\,\sigma$ constraint on $r$ if the true value of $r$ is $5\times 10^{-3}$.} In the last case, we use $E$-mode data obtained by combining \LB\ and \S4.}
 \label{tab:sigmar}
\end{table}

To assess the importance of delensing on constraining $r$, we here compute the expected constraint on $r$. 
We consider the following simple forecast based on the direct likelihood method which has been used in, e.g., Refs.~\cite{B1,BKIX,Namikawa:2020:biref}, where the posterior distribution is evaluated from the simulation without assuming a specific likelihood function. 

We first construct an estimator for $r$. At a single multipole, the tensor-to-scalar ratio, which maximizes the full-sky Gaussian likelihood function of $B$-mode multipoles, is obtained by, e.g., differentiating Eq.~(3) of Ref.~\cite{Katayama:2011eh} in terms of $r$: 
\al{
    \hat{r}_\l = \frac{\hat{C}_\l-C^{r=0,\rm obs}_\l}{C^{\rm IGW}_\l} 
    \,. 
}
Here, $\hat{C}_\l$ is the observed $B$-mode power spectrum, $C^{r=0,\rm obs}_\l$ is a model for the observed power spectrum without the IGW signal, and $C^{\rm IGW}_\l$ is the IGW-induced $B$-mode power spectrum with $r=1$. In our simulation, we adopt the above equation for the binned spectrum and obtain an estimate of $\hat{r}_b$ at each multipole bin. We then combine $\hat{r}_b$ to estimate the single number for $r$ using a weighted mean over multipole bins:
\al{
	\hat{r} = \frac{\sum_b a_b \hat{r}_b}{\sum_b a_b} 
	\,. \label{Eq:estr}
}
The weights, $a_b$, are taken from the covariance of $\hat{r}_b$ obtained from simulation as
\al{
	a_b = \sum_{b'} {\bm {\mathrm{Cov}}} _{bb'}^{-1} 
	\,.
}
The binned angular power spectra, $C^{r=0,\rm obs}_b$ and $C^{\rm IGW}_b$, are both computed from simulations by averaging over realizations. 
In the direct likelihood method \cite{B1}, for a given $r$, we compute the distribution of $\hat{r}$ and obtain the probability $P(\hat{r}|r)$ and then obtain the posterior distribution by assuming $P(r|\hat{r})\propto P(\hat{r}|r)$. 
\tn{
The estimator, $\hat{r}$, is unbiased, and we set $\hat{r}=r_0$ where $r_0$ is the true value assumed in this paper. We use the posterior distribution, $P(r|\hat{r}=r_0)$, to evaluate the expected $1\,\sigma$ error on $r$. 
In our forecast, we choose a nonzero true value, $r_0=5\times 10^{-3}$, to highlight delensing as we described in the introduction. 
%In this case, the recombination bump, where delensing is important, is still not dominated by the cosmic variance from IGWs. 
%The results are also robust against possible other concerns, such as instrumental systematic effects and foreground contaminations, which would lead to uncertainties on $r$ well below $10^{-3}$ \cite{LiteBIRD:PTEP}. 
}
%For $r_0=0$, the $1\,\sigma$ upper bound on $r$, $\delta r$, is defined as the value covering the $68\%$ area of the posterior as defined in Eq.~(20) of \pteppaper. 
%For $r_0=5\times 10^{-3}$, 
We find that the posterior distribution is close to a Gaussian distribution and compute the $1\,\sigma$ statistical error on $r$, $\sigma(r)$, by fitting the posterior distribution with a Gaussian function. 

Figure \ref{fig:posterior} shows the posterior distribution obtained from the direct likelihood approach. We compare the results with and without delensing. For all the delensing cases, we use the lensing map internally reconstructed from \LB\ data. 
Table \ref{tab:sigmar} shows 
\tn{
the expected $1\,\sigma$ statistical uncertainty on $r$ if $r_0=5\times 10^{-3}$. 
}
The $r$ constraint does not improve much by using the internally reconstructed lensing map. Adding either the CIB or galaxies shows a similar improvement on the $r$ constraint. By combining available mass tracers of the CIB, galaxies, and the internally reconstructed lensing map, the delensing improves the constraint by approximately $15\%$ by combining \LB\ and external tracers. If we further combine with \S4\ data, the constraint on $r$ improves by $\sim 20\%$. The results are almost insensitive to the number of multipole bins if the number chosen is around $15$. 
%\textcolor{red}{We note that $\delta r$ is somewhat higher than the value shown in \pteppaper. This could be due to the difference between the forecast methods. Our estimate is fully simulation-based while \pteppaper\ made several simplifications to derive the constraint, e.g., no-correlation between multipoles $\l$ and using the idealistic likelihood. On the other hand, our method would not be optimal to constrain $r$ since our estimate does not include a spatial variation of the residual Galactic foreground noise. Thus, the constraint shown in this paper would not be the best precision of $r$ that \LB\ can achieve. We leave optimization for constraining $r$ as well as a quantitative investigation for the discrepancy between the simulation-based estimate and \pteppaper\ result for future work.}

\section{Summary and Discussion} \label{sec:summary}

As a further investigation of delensing for \LB\ from \pteppaper, we have explored the potential of the \LB\ delensing using multiple tracers, including the CIB from the \Planck\ GNILC method, galaxy spatial distributions from \Euclid\ and LSST tomographic data, and high-resolution CMB data from \S4. We included some realism, such as the survey window for each data set, a realistic \LB\ lensing map, and \LB\ $E$-mode polarization after component separation. We developed algorithms to combine multiple tracers optimally and to combine $E$-mode polarization from \LB\ and \S4, and we constructed our lensing $B$-mode template from the optimally combined lensing mass map and the $E$-mode maps. We also used realistic component-separated $B$-mode polarization obtained by \pteppaper\ to estimate the expected constraint on $r$. 
We found that delensing can improve the constraint on $r$ by around $15\%$ by combining external data from the CIB and galaxies. The improvement becomes $20\%$ when further adding \S4. If the residual foreground noise is reduced, the importance of delensing is further increased. 

In our forecast, we have ignored the uncertainties in the mass tracers. Multiple studies have already investigated the impact of these uncertainties on the estimate of $r$ \cite{Sherwin:2015,BaleatoLizancos:2020:cibdelens,Namikawa:2021:SO-delens,BaleatoLizancos:2022:extgal}. 
% multiplicative bias
A multiplicative bias in mass tracers propagates into the lensing $B$-mode template and modifies its shape. However, the spectral shape is close to a white noise spectrum, similar to the lensing $B$-mode power spectrum. Thus, the bias in $r$ from the uncertainty of the mass tracer would be absorbed into a nuisance parameter such as the amplitude parameter of lensing, $A_{\rm lens}$. 
% extragalactic sources
Additive biases to the mass tracer could also lead to a bias in the lensing $B$-mode template. For example, the internally reconstructed CMB lensing map from temperature could contain extragalactic sources \cite{BaleatoLizancos:2022:extgal}. Such contributions would be mitigated by future polarization-based reconstruction with several methods \cite{Namikawa:2012:bhe,Namikawa:2013:bhepol,Schaan:2018,Sailer:2020:profile,Sailer:2022:bhepol}. 
% CIB residual FGs
In CIB delensing, the residual foregrounds in the CIB could lead to a bias in the delensed $B$-mode power spectrum through higher-order statistics, e.g., $\ave{BEI}_{\rm c}$ and $\ave{EIEI}_{\rm c}$ where $c$ denotes the connected part of the correlation. Ref.~\cite{BaleatoLizancos:2020:cibdelens} showed that, for SO, the bias is negligible for the constraint on $r$. For \LB, the map has a region where the dust contributions are significant compared to the SO patch assumed in Ref.~\cite{BaleatoLizancos:2020:cibdelens}. However, Galactic foreground cleaning of polarization can substantially reduce the bias. Also, multi-frequency data would improve the accuracy of the component separation of both $E$-mode polarization and the CIB map. We expect this bias to be insignificant, but a quantitative analysis is important; however, we leave it for our future work. 

% inhomogeneous noise, instrumental systematics
As an initial investigation of multitracer delensing for \LB\, we have focused on improving constraints on $r$ by adding external tracers and have ignored inhomogeneous noise due to the scan pattern of \LB, as well as any instrumental systematics in our analysis. We also did not include the inhomogeneity of the delensing efficiency in the map. 
Inhomogeneous noise will be accounted for by simply implementing the inhomogeneous noise covariance into the filtering of the observed CMB anisotropies. 
For the instrumental systematics, Ref.~\cite{Nagata:2021} recently performed a preliminary analysis of the impact of instrumental systematics on delensing using a map-based simulation. The results show that the instrumental systematics in the lensing $B$-mode template do not significantly bias the delensed $B$-mode power spectrum as long as the systematics are well suppressed in the measured $B$-mode power spectrum.
The inhomogeneity of the delensing efficiency would also be included, similar to the inhomogeneous noise, by modifying the pixel covariance matrix, which would further improve sensitivity to $r$. 
A more detailed analysis of these effects will be addressed in future work.

%//////////////////////////////////////////////////%
% BACK MATTER 
%//////////////////////////////////////////////////%

% Acknowledgments %
\section*{Acknowledgments}
% =====================================================================================
% LiteBIRD Standard Acknowledgements, long format
% =====================================================================================
%
% Japan; point-of-contact: Masashi Hazumi <masashi.hazumi@kek.jp>
This work is supported in Japan by ISAS/JAXA for Pre-Phase A2 studies, by the acceleration program of JAXA research and development directorate, by the World Premier International Research Center Initiative (WPI) of MEXT, by the JSPS Core-to-Core Program of A. Advanced Research Networks, and by JSPS KAKENHI Grant Numbers JP15H05891, JP17H01115, and JP17H01125.
% Canada point-of-contact: Matt Dobbs <Matt.Dobbs@mcgill.ca>
The Canadian contribution is supported by the Canadian Space Agency.
% France; point-of-contact: Ludovic Montier  <ludovic.montier@irap.omp.eu>
The French \textit{LiteBIRD} phase A contribution is supported by the Centre National d’Etudes Spatiale (CNES), by the Centre National de la Recherche Scientifique (CNRS), and by the Commissariat à l’Energie Atomique (CEA).
% Germany; point-of-contact: Eiichiro Komatau <komatsu@MPA-Garching.MPG.DE>
The German participation in \textit{LiteBIRD} is supported in part by the Excellence Cluster ORIGINS, which is funded by the Deutsche Forschungsgemeinschaft (DFG, German Research Foundation) under Germany’s Excellence Strategy (Grant No. EXC-2094 - 390783311).
% Italy; point-of-contact: Paolo Natoli <ntlpla@unife.it>
The Italian \textit{LiteBIRD} phase A contribution is supported by the Italian Space Agency (ASI Grants No. 2020-9-HH.0 and 2016-24-H.1-2018), the National Institute for Nuclear Physics (INFN) and the National Institute for Astrophysics (INAF).
% Norway; point-of-contact: Hans Kristian Eriksen <h.k.k.eriksen@astro.uio.no>
Norwegian participation in \textit{LiteBIRD} is supported by the Research Council of Norway (Grant No. 263011) and has received funding from the European Research Council (ERC) under the Horizon 2020 Research and Innovation Programme (Grant agreement No. 772253 and 819478).
% Spain; point-of-contact: Enrique Martinez-Gonzalez <martinez@ifca.unican.es>
The Spanish \textit{LiteBIRD} phase A contribution is supported by the Spanish Agencia Estatal de Investigación (AEI), project refs. PID2019-110610RB-C21,  PID2020-120514GB-I00, ProID2020010108 and ICTP20210008.
% Sweden; point-of-contact: "NAME" "EMAIL ADDRESS"
Funds that support contributions from Sweden come from the Swedish National Space Agency (SNSA/Rymdstyrelsen) and the Swedish Research Council (Reg. no. 2019-03959).
% UK; point-of-contact: Erminia Calabrese <calabrese.erminia@gmail.com>
% To be included.
% USA; point-of-contact: Adrian Lee <Adrian.Lee@berkeley.edu>
The US contribution is supported by NASA grant no. 80NSSC18K0132.
%
% =====================================================================================
% Grants
We also acknowledge support from JSPS KAKENHI Grant No. JP20H05859 and No. JP22K03682, and H2020-MSCA-RISE- 2020 European grant (Marie Sklodowska-Curie Research and Innovation Staff Exchange). 
This research used resources of the National Energy Research Scientific Computing Center (NERSC), a U.S. Department of Energy Office of Science User Facility located at Lawrence Berkeley National Laboratory.

% Appendix %
\appendix

% References %
\bibliographystyle{JHEP}
\bibliography{cite}

\end{document}